\documentclass[bibyear]{aa} % if the references are not structured 

\usepackage{graphicx}
\usepackage{txfonts}
\begin{document}
   \title{A case against an X-shaped structure in the Milky Way young bulge}

   \subtitle{}

   \author{Mart\'\i n L\'opez-Corredoira \inst{1,2}}
\institute{
$^1$ Instituto de Astrof\'\i sica de Canarias,
E-38205 La Laguna, Tenerife, Spain\\
$^2$ Departamento de Astrof\'\i sica, Universidad de La Laguna,
E-38206 La Laguna, Tenerife, Spain}

\offprints{martinlc@iac.es}
\titlerunning{A case against an X-shaped yound bulge}
\authorrunning{L\'opez-Corredoira}

   \date{Received xxxx; accepted xxxx}

% \abstract{}{}{}{}{} 
% 5 {} token are mandatory
 
  \abstract
  % context heading (optional)
    {A number of recent papers have claimed the discovery of an X-shape structure in the bulge of our Galaxy in the population of the red clumps.}  
  % {} leave it empty if necessary  
   % aims heading (mandatory)
  {We endeavor to analyze the stellar density of bulge stars in the same regions using a different stellar population that is characteristic of the young bulge ($\lesssim 5$ Gyr). Particularly, we use F0-F5 main-sequence stars with distances derived through photometric parallax.}
   % methods heading (mandatory)
  {We extract these stars from extinction-corrected color-magnitude diagrams in the near-infrared of VISTA-VVV data in some bulge regions and calculate the densities along the line of sight. We take the uncertaintity in the photometric parallax and the contamination of other sources into account, and we see that these errors do not avoid the detection of a possible double peak along some lines of sight as expected for a X-shape bulge if it existed.}
   % results heading (mandatory)
  {Only a single peak in the density distribution along the line of sight is observed, so apparently there is no X-shape structure for this population of stars. Nonetheless, the effects of the dispersion of absolute magnitudes in the selected population might be an alternative explanation, although in principle these effects are insufficient to explain this lack of double peak according to our calculations.}
   % conclusions heading (optional), leave it empty if necessary 
  {The results of the present paper do not demonstrate that previous claims of X-shaped bulge using only red clump stars are incorrect, but there are apparently some puzzling questions if we want to maintain the validity of both the red-clump results and the results of this paper.}


   \maketitle
%
%__________________________________________________________outer_new.tex______

\section{Introduction}

A number of recent papers have claimed the discovery of an X-shape structure in the
bulge of our Galaxy (Nataf et al. 2010, 2015; McWilliam \& Zoccali 2010; Saito et al.
2011; Wegg \& Gerhard 2013). All of these studies base their conclusions on the analysis of red
clump star counts along different lines of sight in the region
$|\ell |\lesssim 10 ^\circ $, $5^\circ \lesssim |b|\lesssim 10^\circ $. 
In this region, these studies find a double peak in
the star counts along the line of sight, which they associate with a double peak in the density distribution and hence a structure of X-shape.
This split of red clump is evident for stars with [Fe/H]$>-0.5$ (Ness et al. 2012, Rojas-Arriagada et al. 2014). But this implication is not straightforward since the red clumps stars do not have a
unique narrow peak in its luminosity function.
As a matter of fact, the
distance between the two density peaks in the putative X-shape structure corresponds to a distance modulus of 0.5-0.7 mag (Wegg \& Gerhard 2013, Fig. 6), where distance between the two peaks increases with height from the plane. This is approximately the difference of the magnitudes of the two peaks of the luminosity function in the luminosity function (Girardi 1999; Bovy et al. 2014; Wegg \& Gerhard 2013, Fig. 5; Nataf et al. 2015, Fig. 4).
Also, Lee et al. (2015) believes the X-shape may be wrong
because it is very likely that the double peak is a manifestation of multiple populations: the helium-enhanced second-generation stars placed on the bright red clump, which is about 0.5 mag brighter than the normal red clump that originated from first-generation stars. 
Gonz\'alez et al. (2015), however, counter that the scenario proposed by Lee et al. (2015) is not possible. Anyway, this coincidence of a difference of about 0.5 mag may cast some doubt on the validity of the results obtained only through red clump stars.

Here we endeavor to analyze the stellar density of bulge stars in the same regions using a different stellar population with an age that is typical of this putative structure. Particularly, we use F0V-F5V stars with distances derived through photometric parallax. 

\section{Near-infrared photometric data of VISTA-VVV}

The VISTA variables in the V\'\i a L\'actea (VVV) is an European Southern Observatory (ESO) public survey with the 4.1 m VISTA (Visible and Infrared Survey Telescope for Astronomy) telescope at Cerro Paranal (Minniti et al. 2010; Saito et al. 2012). This telescope performs observations in the Z, Y, J, H, and Ks near-infrared (near-IR) bands toward the Galactic bulge and part of the disk, covering a total area of 562 square degrees. 
In this work, we use stars with magnitudes of the data from Data Release 2 (DR2).
We take the default aperture corrected magnitudes with 2.0 arcsec diameter.

We use the photometry in filters $J$ and $H$. These filters are suitable given the low extinction even in
Galactic bulge regions. We select some regions within 
$|\ell |\le 10 ^\circ $, $-10^\circ \le b\le -6^\circ $ of VISTA-VVV survey;
VISTA-VVV does not cover the positive latitudes over +5 degrees.
Given that the feature we want to observe, a single or double peak in the density along the line
of sight, would be repeated in the different directions, 
we select a few representative regions with lower extinction.
We select circular regions of radius 0.5$^\circ $ 
centered in Galactic coordinates $(\ell , b)$ with $b=-6.5^\circ , -7.5^\circ , -8.5^\circ , -9.5^\circ $
and $\ell $ such that reddening is minimum in Schlegel et al. (1998) extinction maps; we select first within $2^\circ <|\ell |<10^\circ $ and second within $|\ell |\le 2^\circ $. The selected regions are given in Table \ref{Tab:regions}.

We adopt the approximation that in these off-plane regions the Schlegel et al. (1998) cumulative extinction
applies to all of the stars with heliocentric distance $r>4$ kpc, which corresponds to a height below
the Galactic plane that is larger than 450 pc, where we can consider the amount of dust negligible. For lower $r$ stars, the extinction would be lower, but we do not use the density for stars with $r\le 4$ kpc. We assume extinction ratios $A_H=0.51\times E(B-V)$, $E(J-H)=0.29\times E(B-V)$ (Schlafly \& Finkbeiner 2011).
We also neglect the contamination of galaxies, whose density is much lower than the density of stars within the present ranges of magnitudes and coordinates.

\begin{table}
\caption{Fields of VISTA-VVV used for the analysis. All regions are circles of angular radii equal to 0.5 deg centered in the given coordinates. Extinction $A_H$ calculated from Schlegel et al. (1998).}
\begin{center}
\begin{tabular}{cccc}
Gal. long., lat. (J2000)  & $\langle A_H\rangle$ &  \\ \hline
2.31$^\circ $, -6.50$^\circ $ & 0.136 \\ 
-5.23$^\circ $, -7.50$^\circ $ & 0.107 \\
-5.54$^\circ $, -8.50$^\circ $ & 0.073 \\
-6.16$^\circ $, -9.50$^\circ $ & 0.063 \\ \hline
\label{Tab:fields}
2.00$^\circ $, -6.50$^\circ $ & 0.139 \\ 
-1.24$^\circ $, -7.50$^\circ $ & 0.122 \\
-1.24$^\circ $, -8.50$^\circ $ & 0.083 \\
-1.28$^\circ $, -9.50$^\circ $ & 0.071 \\ \hline
\end{tabular}
\end{center}
\label{Tab:regions}
\end{table}

\section{Method of stellar density determination}

The simplest method of determining the stellar density along a 
line of sight in the disk is by isolating a group of stars with the same color and absolute magnitude $M$ within a extinction-corrected color magnitude diagram (CMD). This allows the luminosity function to
be replaced by a constant in the stellar statistics equation and the differential star counts for each line of sight, $A(m)\equiv \frac{dN(m)}{dm}$, can be immediately converted into density $\rho (r)$,

\begin{equation}
\label{diffsc}
\rho [r(m)]=\frac{5}{\ln 10}\frac{1}{\omega \,r(m)^3}A(m)
,\end{equation}\[
r(m)=10^{[m-M+5]/5}
,\]
where $\omega $ is the area of the solid angle in radians and $r$ is the
distance in parsecs.

For this work we select the sources between F0V and F5V. 
They are bright enough to be observed in the bulge fields and their typical ages are lower than 5 Gyr.
Earlier types (O, B, A) would not be as suitable as these sources, since they would belong 
to a younger population and, moreover, they have a large variation of absolute magnitude with
color. Later types (late F, G, K, M) are faint to be observed at the most distant part of the bulge and, moreover, they would have a high giant contamination for the redder cases.

\subsection{Ages}

The red clumps are a population that extends over a wide range of ages and it is difficult to determine their ages since their infrared colors are almost independent of the age. 
We know that a metal-poor old population belongs to a classical bulge
(Babusiaux et al. 2010; D\'ek\'any et al. 2013; Gonz\'alez et al. 2015), and the putative X-shaped pseudo-bulge, if it exists, has an
age range that depends on the orbital families making the X, and also on the formative history of the galaxy (Athanassoula 2016).

Zoccali et al. (2003) claim that no trace is found for any star younger than 10 Gyr. They base their conclusion in the analysis of optical and near-IR CMDs once they are decontaminated from foreground disk stars. This result contradicts other papers that reveal the existence of younger stars in the bulge (e.g., Cole \& Weinberg 2002; Hill et al. 2011; Bensby et al. 2011, 2013; Nataf 2015; Catchpole et al. 2016). The problem in Zoccali et al. (2010) may be the interpretation of diagrams, such as their Fig. 25, as a total absence of 3-5 Gyr stars; 
there is a gap of F0-F5V stars at $J-H=0.2$, $J=16.5$($H=16.3$) once the disk is substracted, but this is because that is in the outskirts of the bulge in which the heliocentric distance is 6 kpc. The bulge becomes important at $H=17.0$ ($J=17.2$), where the heliocentric distance is 8 kpc, and with this magnitude at $J-H=0.2$, we find many stars both in Fig. 25 of Zoccali et al. (2003) and in our Fig. \ref{Fig:CMD}. The gap affects turn-off stars more than main-sequence stars of 3-5 Gyr stars, and there are few of these stars in Zoccali et al. CMDs simply because they are much less abundant than other populations (giants or dwarfs) and their covered area is very small.

Clarkson et al. (2011) constrain the number of young stars ($<5$ Gyr) to be less than 
3.4\% of the total number of stars in the bulge. 
Even if this result were correct, it does not contradict our results.
As a matter of fact, we obtain a maximum density of $\sim 10^{-3}$ star/pc$^3$ of F0-F5V stars in the line of sight at $\ell=2.3^\circ $, $b=-6.5^\circ $ (Fig. \ref{Fig:dens}/top/left), whereas the total maximum density of stars (for instance, with the boxy bulge of L\'opez-Corredoira et al. 2005) in this line of sight is 0.4 star/pc$^3$, which means that observed F0-F5V stars are only around 1/400 of the total number of bulge stars. There are other stellar types with ages lower than 5 Gyr, for instance, main-sequence stars with types O, A, B, but their number is lower.
In any case, our analysis of CMDs, such as the one shown in our Fig. \ref{Fig:CMD},
clarifies that there is a significant amount of stars with $J-H\approx 0.2$ and $H$ between 16.5 and 18.0 and this cannot be due to foreground disk stars because these colors indicate that they should be F0-F5V and their apparent magnitudes should be much lower. 
Blue stragglers in the bulge (Clarkson et al. 2011) of the same color are brighter than F0-F5V stars and they should be at $M_H$ around 16.5 (see Fig. 25 of Zoccali et al. 2003), which supposes a negligible contamination at regions away from the bulge ($r<6$ kpc).
The only possible ways to posit a possible contamination is that their intrinsic color is wrong and, therefore, they are not F0-F5V stars, or that the theoretical predictions of their color were wrong.

\subsection{Magnitudes and completeness}

Covey et al. (2007) gives the magnitudes of any stellar type, which we use here. 
Covey et al. (2007), however, give 2MASS magnitudes (Vega calibrated) 
and there is some small difference with VISTA-VVV magnitudes (also Vega calibrated).
Using the Hewett et al. (2006) formulas to convert 2MASS into WFCAM(UKIDSS) magnitudes, and using 
the relations to convert WFCAM into VISTA magnitudes \footnote{Available at: 
http://casa.ast.cam.ac.uk/surveys-projects/vista/technical/photometric-properties}, we derive
for main-sequence stars as follows:
\begin{equation}
J_{VISTA}=J_{2MASS}-0.084(J_{2MASS}-H_{2MASS})
\end{equation}\[
\ \ \ \ \ \ -0.021(H_{2MASS}-K_{s,2MASS})+0.005
\]\begin{equation}
J_{VISTA}-H_{VISTA}=0.940(J_{2MASS}-H_{2MASS})
\end{equation}\[
\ \ \ \ \ \ -0.071(H_{2MASS}-K_{s,2MASS})+0.024
\]

\begin{figure}
\vspace{1cm}
\centering
\includegraphics[width=8.5cm]{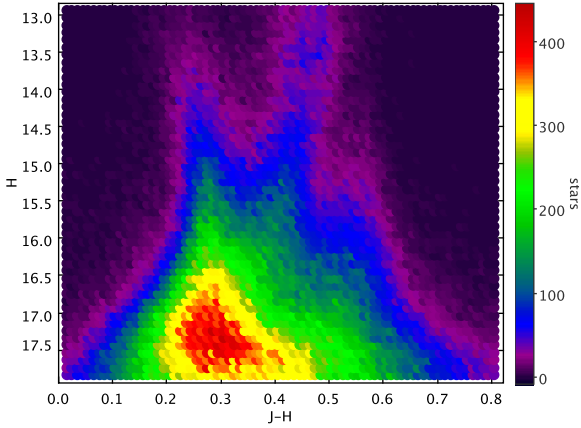}
\caption{Example of extinction-corrected CMD in the region $\ell =-6.2^\circ $, $b=9.5^\circ$. The plot represents the number of stars in bins of 0.01 mag in color and 0.1 mag in $H$ magnitude.
The extinction was assumed to be the same for all stars: the total cumulative one along the line of sight 
given by Schlegel et al. (1998); this is not correct for the nearby sources but, since we analyze only sources with $r>4$ kpc, the use of this CMD is here appropriate.
The selected F0-F5V stars are within $0.15<(J-H)<0.23$.}
\label{Fig:CMD}
\end{figure}

With these transformations and using Table 3 of Covey et al. (2007), for solar metallicity
we obtain for VISTA filters that the F0V-F5V stars have a range of absolute magnitudes $M_H$=2.25 to 2.56; we assume an average $\langle M_H\rangle=2.40$.
The range of intrinsic colors of these stars would be $J-H$ between 0.06 and 0.23, 
Values of $J-H\le 0.14$ are also present in some A dwarfs
so we restrict $J-H$ between 0.15 and 0.23 to minimize their contamination. 
In Fig. \ref{Fig:CMD}, we give an example of how we select these stars.

\begin{figure}
\vspace{1cm}
\centering
\includegraphics[width=8.5cm]{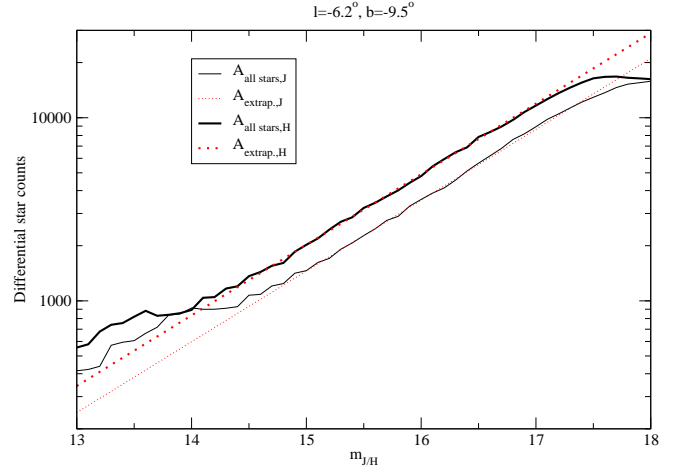}
\caption{Example in the region $\ell =-6.2^\circ $, $b=9.5^\circ $ 
of VISTA-VVV counts of all stars fitting in the range 15.0-16.5 in J and H, 
and its extrapolation to other magnitudes.}
\label{Fig:fitcompl}
\end{figure}

The given absolute magnitudes make the sources 
approximately $m_H\lesssim 17.8+A_H$ within 12 kpc, where $A_H$ 
is the extinction in the filter $H$.
The completeness limit of our survey is around this limit. Nonetheless,
regions with lower $b$ toward the Galactic center are more crowded so 
the completeness limit is lower.
We do not think it matters if we lose some sources because we investigated the shape of the density along the lines of sight and a small number of unobserved stars would only make the  amplitude of the density smaller, but it would not modify its shape. In any case, we also estimate the effect of incompleteness in the counts $A(m)$ by assuming the following correction based on an extrapolation of the counts from the range of magnitudes that we know are complete:
\begin{equation}
\label{corrcomp}
A_{corr.,F0-5V}(m_H)=A_{F0-5V}(m_H)
\end{equation}\[
\ \ \ \ \ \times {\rm Max}\left(\frac{A_{extrap.,J}[m_H+(J-H)_0]}{A_{{\rm all\ stars},J}[m_H+(J-H)_0]},
\frac{A_{extrap.,H}(m_H)}{A_{{\rm all\ stars},H}(m_H)}\right)
,\]
where $(J-H)_0=0.19$ is the average color of the F0-F5V population; and $A_{extrap.,J/H}(m_{J/H})$ is the extrapolation of the fit to the counts of all stars between $15.0\le m_{J/H}\le 16.5$ for the filter J, H of function type $\log _{10}A(m)=a+b\,m$. The fit is very good in that range $15.0\le m_{J/H}\le 16.5$ (an example is shown in Fig. \ref{Fig:fitcompl}), but the extrapolation may introduce some inaccuracy due to some departure of this function type and
the scattering of colors. 
This correction only gives an estimation of the effect of the incompleteness, providing a more accurate result for the outgoing density an estimate without such a correction. 

\subsection{Contamination due to photometric errors}

A variation of the distribution is derived from the effect of the photometric errors, 
because of, for example, either instrumental noise or confusion of sources.
The effect is twofold: there is a variation in the apparent magnitude $m_H$ of each source
and there is a variation in the color $(J-H)$ of this source. The first effect does not produce
an important change in the density distribution, rather it results in a very small 
shift in the position of the peaks. The second effect is more severe since small variations of colors are related to large variations of absolute magnitudes. And the error in these colors is conspicuous. For instance, in the region $\ell=-5.2^\circ $,
$b=-7.5^\circ $ at distance $r\approx 8$ kpc, the rms of the color due to error in the magnitude
is $\approx 0.08$. This means a significant number of stars that we count within the
range $0.15<(J-H)<0.23$ stars have colors away from this range and, viceversa, that some stars
that truly have $0.15<(J-H)<0.23$ are not counted because they are
placed away from that range with the photometric errors. Since the measured density is proportional to the measured stars counts through Eq. (\ref{diffsc}), the variation of these counts modifies the observed density.
A calculation of this effect is carried out as follows: the observed density with photometric
errors $\rho ^{\rm phot.err.}$ is
related to the density without photometric errors $\rho ^{\rm no\ err.}$ through
\begin{equation}
\label{photerr}
\rho (r)^{\rm phot.err.}=\rho ^{\rm no\ err.}(r)+
\frac{1}{0.08\sqrt{2\pi }\sigma (r)}\int _{0.15}^{0.23}\,\frac{dc_1}{\phi (c_1)}
\end{equation}\[
\times\left[\int _{-\infty}^{0.15}dc_2\,\Delta \rho _{2,1}e^{-\frac{(c_2-c_1)^2}{2\,
\sigma (r)^2}} +\int _{0.23}^{\infty}dc_2\,\Delta \rho _{2,1}e^{-\frac{(c_2-c_1)^2}{2\,
\sigma (r)^2}} \right]
,\]\[
\Delta \rho _{2,1}\equiv \rho ^{\rm no\ err.}(r_2)\left(\frac{r_2}{r_1}\right)^3\phi (c_2)-
\rho ^{\rm no\ err.}(r_1)\phi(c_1)
,\]\[
\phi (c_i)\equiv \phi [M_H(c_i)]; \ \ r_i=r\times 10^{-\frac{(M_H(c_i)-2.40)}{5}}
\] 
where $c_1$ is the observed color $(J-H)$ and $c_2$ is the real color $(J-H)$, $\sigma (r)$ is the
error of $(J-H)$ for stars of our selected stars (with $M_H\approx 2.40$) at distance $r$.
The variable $\phi (M_H)$ is the luminosity function in the H-band for
bulge stars, which we take from a fit of the Zoccali et al. (2003) luminosity function in
the range, where we take $\phi (M_H)=\phi _0\times 3.04^{M_H}$; and $M_H(c)$ 
is the absolute magnitude of main-sequence stars as a function of their
color, which we take from a fit to Covey et al. (2007, table 3) magnitudes and colors, i.e.,
\begin{equation}
\label{mh}
M_H(c)= -0.10636+10.368c+19.23c^2-5.5335c^3
\end{equation}\[
\ \ \ \ \ \ -375.3c^4+830.03c^5-489.99c^6
\]

\begin{figure}
\vspace{1cm}
\centering
\includegraphics[width=8.5cm]{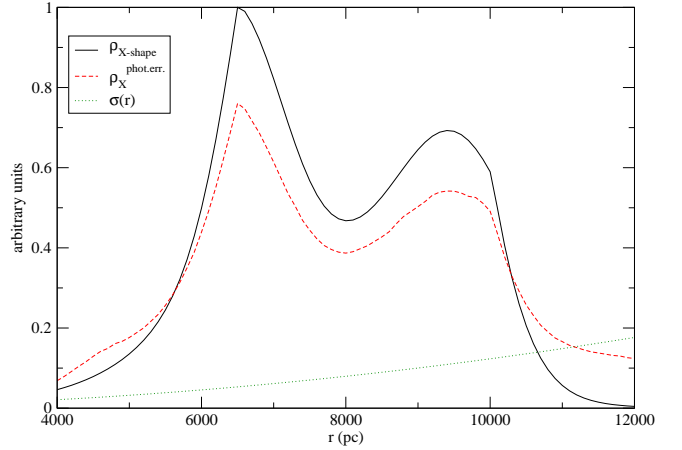}
\caption{Example of the effect of the errors in the photometry in a density distribution
as given by Eq. (\ref{photerr}). The example of $\rho _{\rm X-shape}$ 
is taken from the predictions of the X-shape bulge of Eq. (\protect{\ref{densx}}) 
at $\ell=-5.2^\circ $,
$b=-7.5^\circ $. The rms $\sigma $(r) is the average of errors of H magnitudes (extinction corrected)
in this VISTA-VVV fields
for stars in the field with the corresponding apparent magnitude of distance $r$ and
absolute magnitude $M_H=2.40$.}
\label{Fig:photerr}
\end{figure}

Equation (\ref{photerr}) indicates that the density without corrections
($\rho ^{\rm no\ err.}$) is modified to give the correct density ($\rho ^{\rm phot.\ err.}$) by adding a term that is the integration over all of the values of the measured color range $c_1$ between 0.15 and 0.23 of the exchange of stars due to photometric error with true color less than 0.15 (the first integral within the
square bracket) and those with true color that are larger than 0.23 (the second integral within the square bracket), assuming Gaussian dispersion in the photometric error.
The factor $\Delta \rho _{2,1}$ gives the difference of star density between true color $c_2$ and measured color $c_1$, including the factor proportional to $r^3$ in the counts.
It is the excess/deficit of density multiplied by luminosity function, $\Delta \rho _{2,1}$, which produces the variation of observed density with respect to the real density. 
If we had $\Delta \rho _{2,1}=0\,\forall c_1\ \forall c_2\ \forall r$, we would have no variation of
density $\rho $. Also, for each distance $r$, 
if $\Delta \rho _{2,1}$ had a linear dependence with $c_2$ for a given $c_1$ then, by symmetry,
the excess of the stars that would be gained from one side of the color range would be lost by the same amount on the opposite side. The nonlinear dependence of $\Delta \rho _{2,1}(c_2|c_1)$ results in variations, i.e., in the case when the net gain of stars is different from the total loss. In our case, given that the luminosity function $\phi $ increases strongly with absolute magnitude and $M_H$ is monotonously increasing with $c$ (from Eq. (\ref{mh})), most of the contribution of the variations comes from the integral within $c_2>0.23$, i.e., the exchange of colors between the F0-F5V stars and types later than F5, which are more numerous. In Fig. \ref{Fig:photerr}, we show the transformation of the density for the same region
at $\ell=-5.2^\circ $, $b=-7.5^\circ $. The effect is significant; 
we note however that the peaks are still clearly observed.

\subsection{Dispersion of absolute magnitudes}
\label{.disp}

We also take the effecto of the dispersion of absolute magnitudes into account, i.e., the variation of the absolute magnitudes with respect to the used value of $M_H=2.4$. This has two components: 1) the
variation of the mean $M_H$ due to the variation of the color from $J-H=0.15$ to 0.23 and
2) the dispersion
of absolute magnitudes for a fixed color. The possibility that some of these stars might indeed be binary systems (Siegel et al. 2002) is included within this second component.

The two mentioned variations of absolution magnitude are calculated as follows:
\begin{enumerate}
\item According to Covey et al. (2007), the maximum
variation of average $M_H$ of F0-5V stars with respect to the central value of 2.40 is $\Delta M_H=0.15$ mag. Assuming the same contribution of stars for each color, the smoothing that it produces can be roughly calculated as
\begin{equation}
\label{smooth1}
\rho _{X,smooth-1}(r)\approx 
\int _{r-\Delta r}^{r+\Delta r} dr'\ \left(\frac{r}{r'}\right)^3\rho _{\rm X-shape}(r')
\end{equation}\[
\Delta r=\frac{ln(10)}{5}r\,\Delta M_H
\]

\item Moreover, we should smooth the curve again by performing a convolution with the distribution of absolute
magnitudes around the average value. Assuming a Gaussian distribution of absolute magnitudes,
\begin{equation}
\label{smooth2}
\rho _{X,smooth-2}(r)=\frac{1}{\sqrt{2\pi }\sigma _r}
\int _{0}^{\infty} dr'\ \left(\frac{r}{r'}\right)^3\rho _{\rm X,smooth-1}(r')
e^{-\frac{(r-r')^2}{2\,\sigma _r^2}} 
\end{equation}
\[
\sigma _r=\frac{ln(10)}{5}r\,\sigma_{M_H}
\]

\end{enumerate}

\begin{figure}
\vspace{1cm}
\centering
\includegraphics[width=8.5cm]{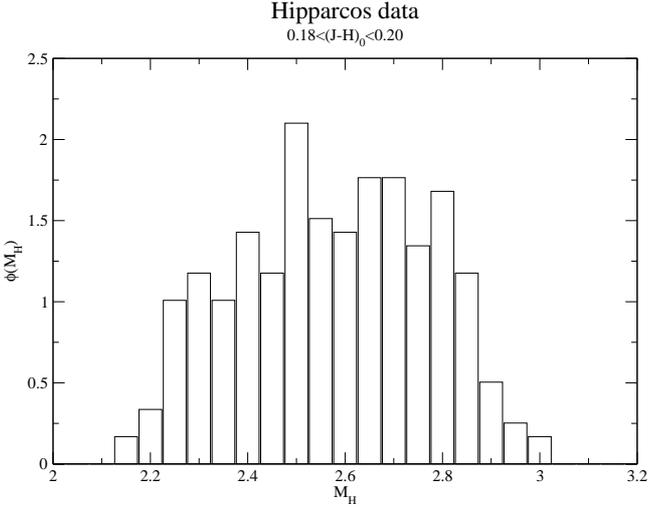}
\caption{Distribution of absolute magnitudes in filter H for extinction-corrected
Hipparcos data of dwarf stars within $0.18<(J-H)_0<0.20$ (taken from Bilir et al. 2008).}
\label{Fig:Hipparcos}
\end{figure}

\begin{figure}
\vspace{1cm}
\centering
\includegraphics[width=8.5cm]{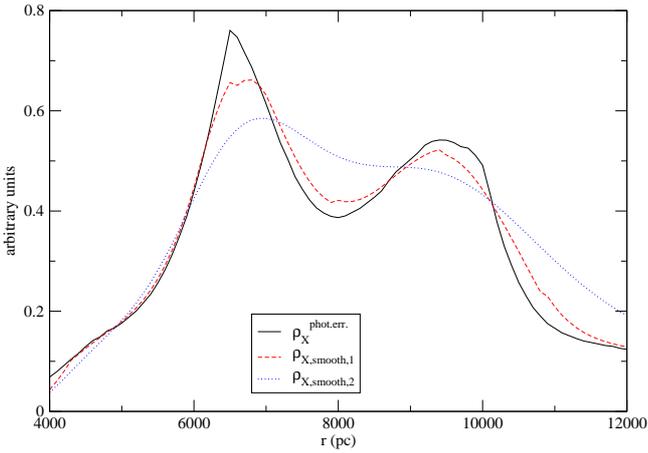}
\caption{Example of the effect of smoothing in a density distribution
due to the dispersion of absolute magnitudes, given by Eqs. (\ref{smooth1}) and
(\ref{smooth2}). The example of
the density distribution with photometric errors correction in Fig. \ref{Fig:photerr}; we set the value of $\sigma _{M_H}=0.19$.}
\label{Fig:smooth}
\end{figure}

In Fig. \ref{Fig:smooth}, the effect of this smoothing is illustrated. 
The example $\rho _{\rm X-shape}$ 
is taken from the density distribution with photometric errors correction in Fig. \ref{Fig:photerr}, and we set the value of $\sigma_{M_H}=0.19$ mag.
We derive this dispersion of 0.19 from Bilir et al. (2008) in two ways.
First, we derive the dispersion in the difference between near-infrared absolute magnitude derived from Hipparcos and
the same absolute-magnitude from Bilir et al. (2008, eq. 4), including color dependence.
Second, we derive the dispersion that we measure in the distribution of Hipparcos data for a given $(J-H)=0.18-0.20$, that is
illustrated in Fig. \ref{Fig:Hipparcos}; the value of the average $M_H$ is 2.57 instead of 2.40, but this does not matter, since this change in the average results in a shift in distance of the counts, but not its smoothing.

The effect of the contamination plus dispersion of magnitudes, given by Eqs. (\ref{photerr}), 
(\ref{smooth1}), and (\ref{smooth2}) is very conspicuous and reduces most of the amplitude of the peaks, but nevertheless the two peaks are distinguished and the distribution is different from a pure single peak distribution.

\subsection{Effect of metallicity variations}

As said previously, we are tracing a population that is younger than 5 Gyr, so no very low metallicity stars are expected in the bulge and the approximation of solar metallicity is acceptable. The average metallicity in the Baade window of the bulge
population is [Fe/H]=-0.11$\pm 0.04$ (Sadler et al. 1996); there is
a gradient of metallicity with the height
(Tiede et al. 1995, Kunder et al. 2012, -0.4 dex/kpc)
or with radius (e.g., according to models by Mart\'\i nez-Valpuesta \& Gerhard 2013: of -0.26 dex/kpc).
A variation of $\frac{dM_V}{d[Fe/H]}\sim 1$ mag dex$^{–1}$ at the blue end (as it is the case of 
F2-F5V stars) is expected (Juri\'c et al. 2008, Fig. 1). The variation of color is negligible.
For instance, from Tables 3 and 5 of Covey et al. (2007), we see that F5V stars 
have the same $g-z=0.27$ either for [Fe/H]=0 or [Fe/H]=-0.30 and the variation for near-infrared colors is expected to
be even smaller given that larger variations of colors are shown toward bluer filters. Therefore,
the color selection should not be affected by the metallicity change, although its absolute magnitude
should be affected. A maximum deviation of -0.5 dex in the
metallicity with respect to the solar value is expected. This deviation is given by the age-metallicity relation  for stars with less than 5 Gyr (Carraro et al. 1998), or derived from a radial gradient
along the major axis of -0.26 dex/kpc (Mart\'\i nez-Valpuesta \& Gerhard 2013) 
with a maximum Galactocentric distance of $\sim 2$ kpc.
This deviation of -0.5 dex would produce an error of the distance 
of the stars of $\lesssim 25$\%.
The shape of the density distribution along the line of sight may also be affected, making
thicker or thinner the peaks, but not removing some of the two possible peaks or producing
new peaks where there were no peaks.
In a putative double-peak structure, because of an X-shaped bulge, each peak may have a small
shift. The most important factor would stem from the
difference of $\Delta z=0.33-0.50$ kpc, using the model given by Eq. (\ref{densx}) for an X-shape bulge, in the used range of Galactic latitudes. This would result in a reduction in the separation of both peaks of 0.13-0.20 mag, which is much lower than the expected separation of peaks of 0.5-0.7 mag., so this effect could not screen the double peak if it were real.

\subsection{Effect of the extinction uncertainties}

Furthermore, we have the uncertainty in the extinction.
We assume that most of the extinction stems from the local disk. 
With extinctions of $\langle A_H\rangle \lesssim 0.15$ mag, 
$E(J-H)=0.57A_H$  and a relative uncertainty of the reddening of 
$\sim 10$\% (Schlegel et al. 1998; M\"ortsell 2013), we derive relative
errors of $\Delta (J-H) \lesssim 0.0086$ for each region; negligible variation of color. 
With a $\frac{dM_H}{d(J-H)}\approx 2.1$ (computed from Covey et al. 2007 table, between F0V and F5V stars), we compute uncertainties in $\Delta M_H\lesssim 0.018$, negligible too.

\section{Results}

\begin{figure*}
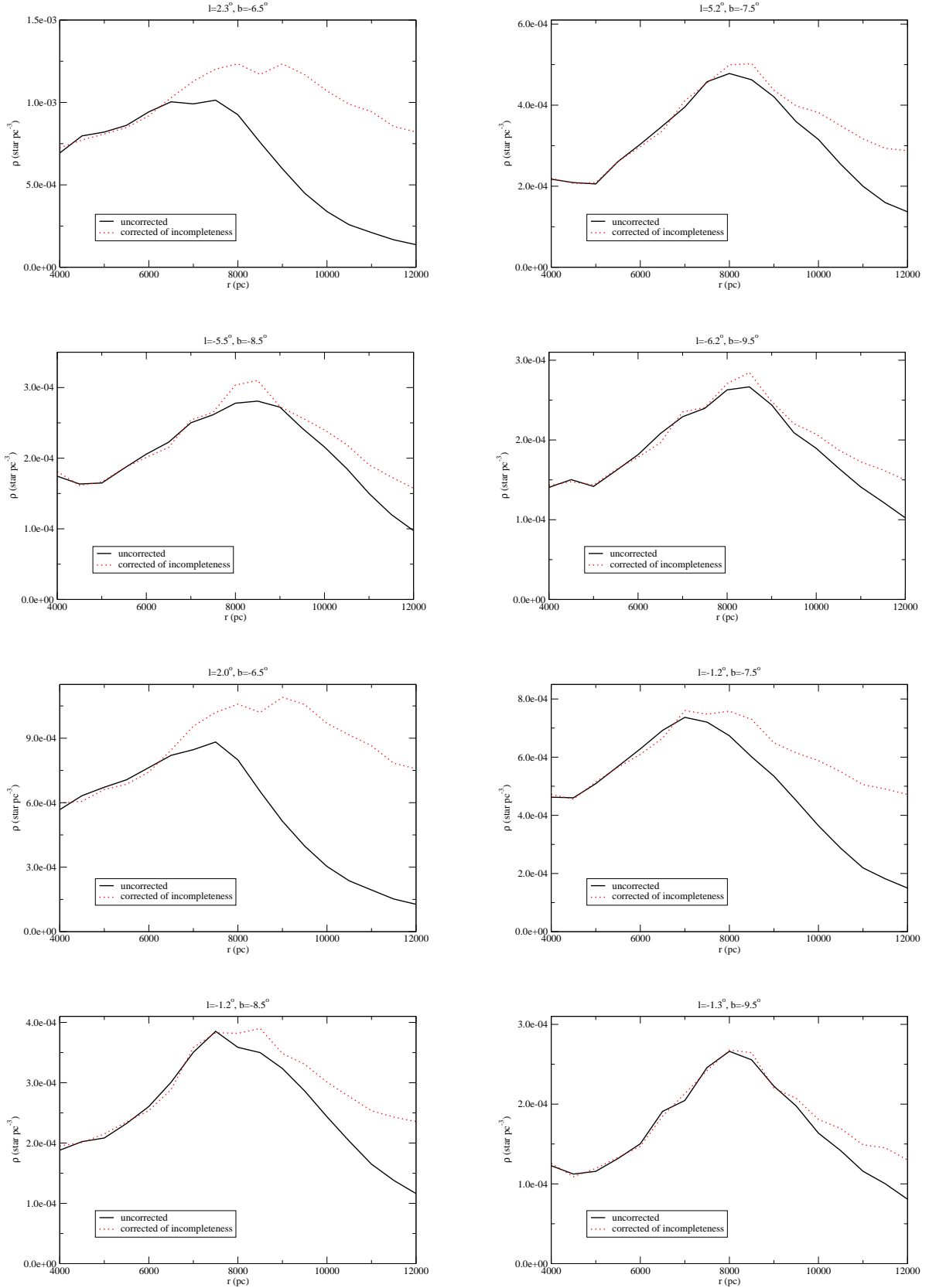

\vspace{.5cm}
\centering
\includegraphics[width=7.2cm]{densl02p3bm6p5.eps}
\hspace{1cm}
\includegraphics[width=7.2cm]{denslm5p2bm7p5.eps}
\\
\vspace{.7cm}
\includegraphics[width=7.2cm]{denslm5p5bm8p5.eps}
\hspace{1cm}
\includegraphics[width=7.2cm]{denslm6p2bm9p5.eps}
\vspace{.7cm}
\\
\includegraphics[width=7.2cm]{densl02p0bm6p5.eps}
\hspace{1cm}
\includegraphics[width=7.2cm]{denslm1p2bm7p5.eps}
\\
\vspace{.7cm}
\includegraphics[width=7.2cm]{denslm1p2bm8p5.eps}
\hspace{1cm}
\includegraphics[width=7.2cm]{denslm1p3bm9p5.eps}
\caption{Density along different lines of sight for F0-F5V stars. The solid line stands for
the density derived from the star counts through Eq. (\protect{\ref{diffsc}})
and the dashed line furthermore includes the estimated correction of incompleteness of 
counts given by Eq. (\protect{\ref{corrcomp}}).}
\label{Fig:dens}
\end{figure*}

\begin{figure}
\vspace{1cm}
\centering
\includegraphics[width=8.5cm]{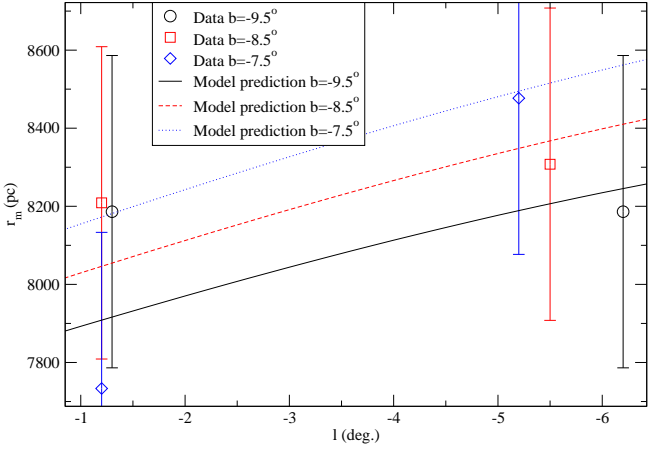}
\caption{Distance of the centroid of the peaks by fitting Gaussian
profiles to the plots of Fig. \protect{\ref{Fig:dens}} in comparison
with the theoretical prediction using Eq. (10) of L\'opez-Corredoira et al. (2007) with elliptical bulge model of L\'opez-Corredoira et al. (2005).}
\label{Fig:fitang}
\end{figure}

The results of the line-of-sight density for the directions given in Table \ref{Tab:regions}
are shown in Fig. \ref{Fig:dens}. All the lines of sight present a single peak associated
with the bulge that is wider and flatter for lower latitudes.
A simple inspection by fitting the centroids of these peaks to Gaussians  shows a distance around
8 kpc, as expected for the center of the bulge. 
If we take the lines of sight with the
six regions with $b\le -7.5^\circ $ (the position of the centroid of the other two lines of sight is too inaccurate because of the large incompleteness corrections) we observe some slightly higher distances at lower Galactic longitude (see Fig. \ref{Fig:fitang}), as expected for a barred structure. In Fig. \ref{Fig:fitang}
we also show a comparison with the expected values in the model of ellipsoidal bulge by L\'opez-Corredoira et al. (2005). The theoretical value
of the maximum density along the line of sight 
does not follow the major axis of the Galactic bulge;
we calculate this value using Eq. (10) of L\'opez-Corredoira et al. (2007, Appendix A). Unfortunately, 
the lines of sight are very few and the error bars of the centroids of the density are not small enough to determine
the angle of the triaxial bulge or other characteristics of its shape accurately, 
which is not the purpose of this paper, 
but at least the data show consistency with what is expected.

%\begin{figure*}
%\vspace{.5cm}
%\centering
%\includegraphics[width=7.2cm]{densl02p3bm6p5.eps}
%\\
%\vspace{.7cm}
%\includegraphics[width=7.2cm]{denslm5p2bm7p5.eps}
%\\
%\vspace{.7cm}
%\includegraphics[width=7.2cm]{denslm5p5bm8p5.eps}
%\\
%\vspace{.7cm}
%\includegraphics[width=7.2cm]{denslm6p2bm9p5.eps}
%\caption{Density along different lines of sight for F0-F5V stars. The solid line stands for the density derived from the star counts through Eq. (\protect{\ref{diffsc}}) and the dashed line furthermore includes the estimated correction of incompleteness of counts given by Eq. (\protect{\ref{corrcomp}}).}
%\label{Fig:dens}
%\end{figure*}
%\begin{figure*}
%\vspace{.5cm}
%\centering
%\includegraphics[width=7.2cm]{densl02p0bm6p5.eps}
%\\
%\vspace{.7cm}
%\includegraphics[width=7.2cm]{denslm1p2bm7p5.eps}
%\\
%\vspace{.7cm}
%\includegraphics[width=7.2cm]{denslm1p2bm8p5.eps}
%\\
%\vspace{.7cm}
%\includegraphics[width=7.2cm]{denslm1p3bm9p5.eps}
%\\
%\vspace{.7cm}
%Fig. \ref{Fig:dens} (cont.)
%\end{figure*}

\subsection{Comparison with models}

We can compare our results with the expected results in an X-shaped bulge or a boxy bulge. 
We evaluate the density $\rho (x,y,z)$, 
where $x$ is the projection along the major axis of the bulge, 
$y$ is the projection perpendicular to the major axis in the plane, and
$z$ is the distance from the plane. We set the angle between the major axis and line Galactic center-Sun
as $\alpha =27^\circ $ (L\'opez-Corredoira et al. 2005; Wegg \& Gerhard 2013). We assume in all cases
a distance to the Galactic centre of $R_0=8000$ pc.

\begin{description}
\item[X-shaped:] we cannot find in the literature a simple analytical expression to evaluate the density. We use a formula that reproduces
approximately the density profiles given by Wegg \& Gerhard (2013, Fig. 12) as follows:
\begin{equation}
\label{densx}
\rho _{{\rm X-shape}}(x,y,z)=\rho _0\times \exp\left(-\frac{s_1}{700\,{\rm pc}}\right)\times \exp \left(-\frac{|z|}{322\,{\rm pc}}\right)
\end{equation}\[
\ \ \ \ \times \left[1+3\times \exp\left(-\left(\frac{s_2}{1000\,{\rm pc}}\right)^2\right)
+3\times \exp\left(-\left(\frac{s_3}{1000\,{\rm pc}}\right)^2\right) \right]
,\]\[
s_1={\rm Max}\left[2100\,{\rm pc},\sqrt{x^2+\left(\frac{y}{0.7}\right)^2}\right]
,\]\[
s_2=\sqrt{(x-1.5z)^2+y^2}
\]\[
s_3=\sqrt{(x+1.5z)^2+y^2}
.\]

\item[Boxy:] we adopt the boxy-bulge model obtained from inversion of stellar star counts by
L\'opez-Corredoira et al. (2005) as follows:
\begin{equation}
\rho _{\rm Boxy}(x,y,z)=\rho _0\times \exp\left(-\frac{\left(x^4+\left(\frac{y}{0.5}\right)^4+\left(\frac{z}{0.4}\right)^4\right)^{1/4}}{740\,{\rm pc}}\right)
\end{equation}

\begin{figure}
\vspace{.5cm}
\centering
\includegraphics[width=9cm]{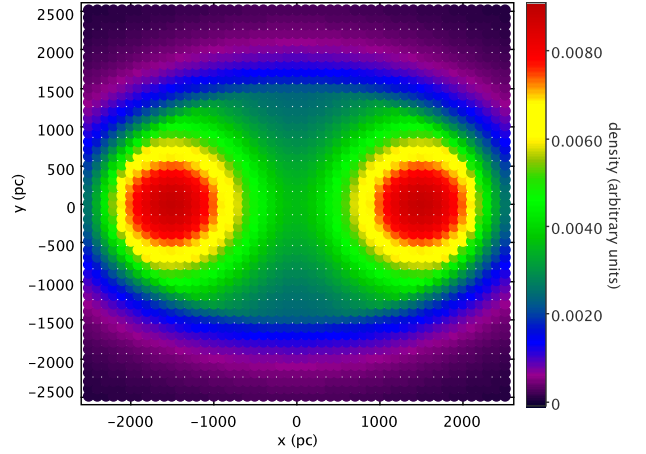}
\\
\vspace{.7cm}
\includegraphics[width=9cm]{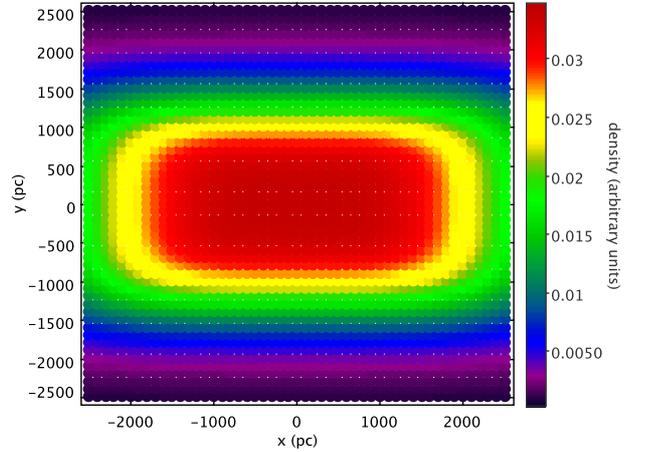}
\caption{Density of the bulge at $z=1000$ pc in the two models: X-shaped (top), boxy (bottom).}
\label{Fig:modz1000}
\end{figure}

\end{description}

The stars studied in this paper constitute a small part of the whole bulge, only around 4\% of the bulge stars. Anyway, recognizing the X-shape in a small number statistics would be possible if it were real because we have a density in the maxima of the peaks of $\gtrsim 3\times 10^{-4}$ star pc$^{-3}$, which is equivalent at the distance of the center of the Galaxy to $\gtrsim  17\,000$ star\,mag$^{-1}$ (using Eq. (\ref{diffsc})). In the plot in Fig. \ref{Fig:dens}, we use a binning of $\Delta r=500$ pc, which is equivalent to $\Delta m=0.14$ mag, and so we have $\gtrsim  2\,300$ stars per bin, which gives a very small Poissonian relative error.

\begin{figure*}
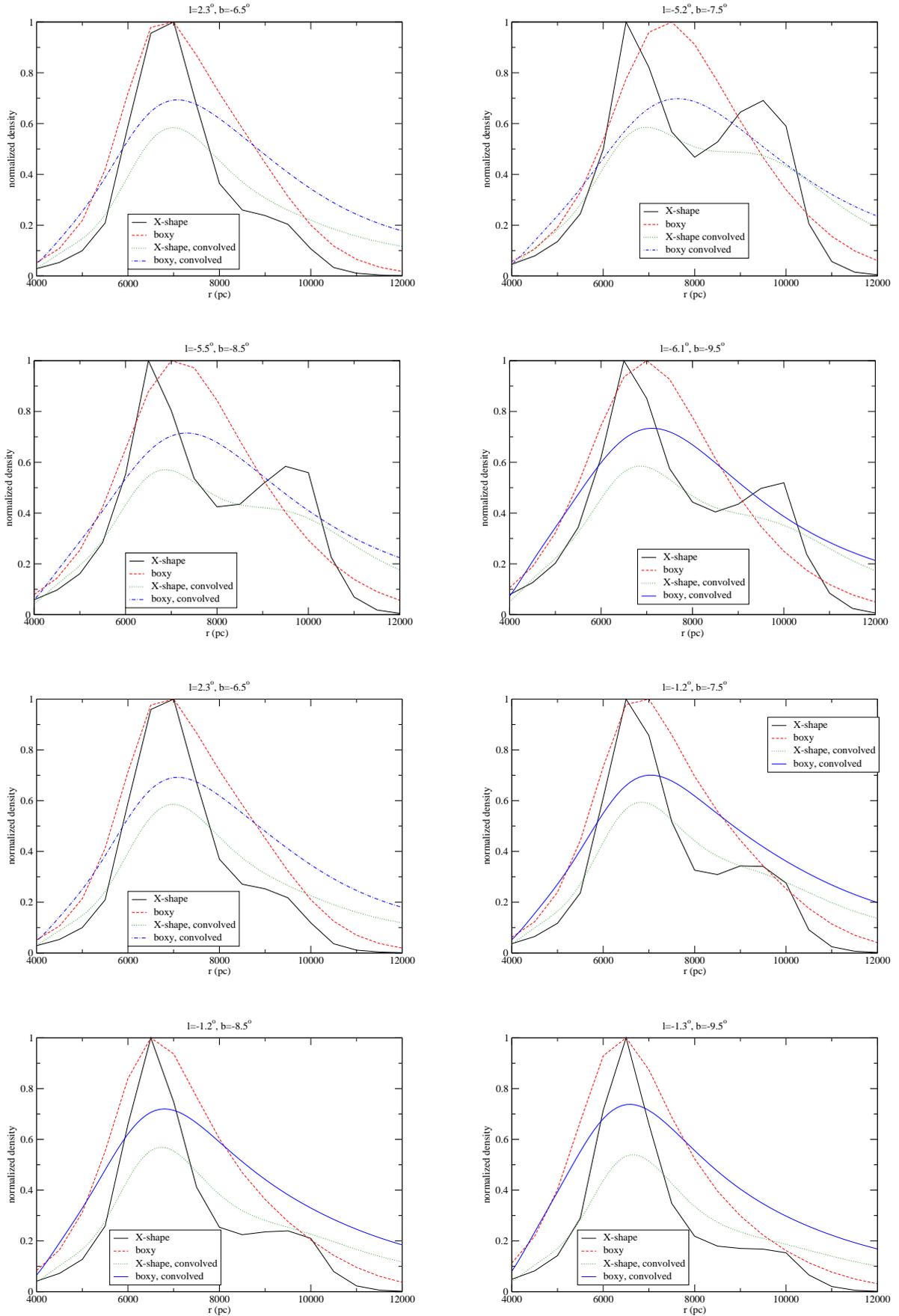

\vspace{.5cm}
\centering
\includegraphics[width=7.2cm]{modl02p3bm6p5.eps}
\hspace{1cm}
\includegraphics[width=7.2cm]{modlm5p2bm7p5.eps}
\\
\vspace{.7cm}
\includegraphics[width=7.2cm]{modlm5p5bm8p5.eps}
\hspace{1cm}
\includegraphics[width=7.2cm]{modlm6p2bm9p5.eps}
\vspace{.7cm}
\\
\includegraphics[width=7.2cm]{modl02p0bm6p5.eps}
\hspace{1cm}
\includegraphics[width=7.2cm]{modlm1p2bm7p5.eps}
\\
\vspace{.7cm}
\includegraphics[width=7.2cm]{modlm1p2bm8p5.eps}
\hspace{1cm}
\includegraphics[width=7.2cm]{modlm1p3bm9p5.eps}
\caption{Model predictions of density of X-shaped and boxy bulge for the same lines of sight
of Fig. \protect{\ref{Fig:dens}}. Only the bulge is represented; the disk is not included.
The convolved curves correspond to the convolution of the density due to photometric errors,
contamination, and dispersion of absolute magnitudes of Eqs. (\ref{photerr}), 
(\ref{smooth1}), and (\ref{smooth2}).}
\label{Fig:mod}
\end{figure*}

In Fig. \ref{Fig:modz1000}, we plot an illustrative example of the density distribution for both models at $z=1$ kpc. The predictions of these two models for the lines of sight of Table \ref{Tab:regions} are given in Fig. \ref{Fig:mod}. We do not include the disk here, only the density of bulge stars is included. We also include the effect of the photometric errors, contamination, and dispersion of absolute magnitudes.

%\begin{figure*}
%\space{.5cm}
%\centering
%\includegraphics[width=7.2cm]{modl02p3bm6p5.eps}
%\\
%\vspace{.7cm}
%\includegraphics[width=7.2cm]{modlm5p2bm7p5.eps}
%\\
%\vspace{.7cm}
%\includegraphics[width=7.2cm]{modlm5p5bm8p5.eps}
%\\
%\vspace{.7cm}
%\includegraphics[width=7.2cm]{modlm6p2bm9p5.eps}
%\caption{Model predictions of density of X-shaped and boxy bulge for the same lines of sight
%of Fig. \protect{\ref{Fig:dens}}. Only the bulge is represented; the disc is not included.
%The convolved curves correspond to the convolution of the density due to photometric errors,
%contamination and dispersion of absolute magnitudes of Eqs. (\ref{photerr}), 
%(\ref{smooth1}) and (\ref{smooth2}).}
%\label{Fig:mod}
%\end{figure*}
%\begin{figure*}
%\vspace{.5cm}
%\centering
%\includegraphics[width=7.2cm]{modl02p0bm6p5.eps}
%\\
%\vspace{.7cm}
%\includegraphics[width=7.2cm]{modlm1p2bm7p5.eps}
%\\
%\vspace{.7cm}
%\includegraphics[width=7.2cm]{modlm1p2bm8p5.eps}
%\\
%\vspace{.7cm}
%\includegraphics[width=7.2cm]{modlm1p3bm9p5.eps}
%\\
%\vspace{.7cm}
%Fig. \ref{Fig:mod} (cont.)
%\end{figure*}

The comparison between Figs. \ref{Fig:dens} and \ref{Fig:mod} points out that the
two peaks predicted by the X-shaped bulge are not reproduced by the observations, whereas the
predictions of the boxy bulge resemble the data better. There are some lines of sight where it is not possible to distinguish the
predictions of both models, but some directions at least show such a distinction, and the models
with one single peak resemble the observations better than the models with two peaks.
A disk model is not included here. This model
would increase the counts in the wings of the single peak, but for the center of the bulge the 
contribution of the disk would be negligible in these off-plane regions.

The distance of the
maximum of the single peak is observed coincident with the prediction, including the convolution.
McWilliam \& Zoccali (2010) use a value of the red clump absolute magnitude, of $M_K=-1.44$, which is fainter
than the usual accepted value at around -1.6 or -1.7. This produces the effect of a higher distance and thereby creates the unique effect of X-shape structure centered in the Galactic center.
Saito et al. (2011) use the same distances derived for the two peaks by McWilliam \& Zoccali (2010), therefore they are implicitly using the same fainter calibration and overestimated distances. Wegg \& Gerhard (2013), however,
appropriately reproduce the double peak around a Galactic center with the usual Galactocentric distance and $M_K=-1.72$.
Anyway, any difference in the assumed absolute magnitude of the red clump simply causes systematic offsets in the distance. This does not undermine the result that is obtained independently by different authors.

\section{Discussion and conclusions}

We do not observe a double peak in the bulge density along the lines of sight within
$|\ell |\lesssim 10 ^\circ $, $-10^\circ \lesssim b\lesssim -6^\circ $ for F0-5V stars.
The dispersion of absolute magnitudes may dilute part of the distinction of these two peaks
but at least in some lines of sight such a distinction should be observed if the two peaks existed,
and this is not the case here.
We tentatively interpret this fact as an absence of X-shape in the young bulge.
D\'ek\'any et al. (2013) or Gran et al. (2016) have also observed that a different tracer, RR Lyrae, does not yield an X-shaped bulge, but they attribute this to a very old population, assuming that the X-shaped bulge is only observed in a relatively younger population.
As a matter of fact, the existence of RR Lyrae stars 
in the bulge reveals to us the existence of stars that are older than 10 Gyr and are a majority (Rich 1993; Zoccali et al. 2003; Clarkson et al. 2011).
A pure pseudo-bulge or pure classical bulge, however, is ruled out by the observations
of the stellar populations (e.g., Babusiaux et al. 2010).
Therefore, if the old population is not X-shaped and the young population is not X-shaped either, one may wonder when this putative X-shape is formed.
If we assumed that the results of this paper are correct and there is a non-X-shaped bulge in the population of $\lesssim 5$ Gyr, and knowing that the older population with low metallicity is also non-X-shaped, it would cast doubt upon the possible existence of a third population of intermediate-old age and metal rich with X-shape. The problem of this scenario is that, apart from the fact that three bulges have never been observed in any galaxy and or in any simulation, there is not a plausible explanation for the formation of a new structure of the bulge after the formation of an X-shaped bulge. While not all of the stars of the X-shaped bulge must be young, one would expect that the stars that are younger than the formation of the X-shaped bulge would have an X-shaped density distribution if that bulge existed. Nonetheless, one may wonder whether there is the possibility that young stars were formed out of gas accreted long after the X-shape was formed by buckling. Also, it is not clear in which way the vertically thick bar component might interact with the thick disk or with the environment. 
The results of the present paper do not demonstrate that previous claims of X-shaped bulge using only red clump stars are incorrect, but there are apparently some puzzling questions if we want to maintain the validity of both the red-clump results and the results of this paper. 

Could the second peak of the analysis of density with red clumps be due to an artefact in the contamination of the luminosity function?
In the double-peak detection of the red clumps, the difference of the distance moduli of both peaks is around the same as the difference of both clump peaks in the luminosity function: a difference of 0.5-0.7 magnitudes, which makes the case suspicious.
This distance between the two red clumps 
slightly decreases toward the Galactic plane (McWilliam \& Zoccali 2010),
but this does not prove anything in favor of an X-shaped structure. This variation of the distance modulus might be due to the vertical metallicity gradient that varies the difference of absolute magnitudes between the two red clumps, although Wegg et al. (2013) find their results of double peak to be consistent with a small metallicity gradient for the red clump, and
therefore this would not account for the decreasing separation. In any case, this is only a tentative explanation for the red clump double peak to be further explored.

A recent claim by Ness \& Lang (2016) to have directly observed
the X-shaped structure in the infrared image of the bulge 
depends on the disk model that is subtracted in the plane region. In the region away from the plane, it looks a peanut shape, that
could be reproduced by other models of the bulge without an X shape. In any case, these putative wings of the X shape are in the tangential direction, and not along the line of sight as observed by the red clump, so this dubious result cannot confirm the conclusion of the 
red clump X-shape either.
A defence of the X-shaped bulge hypothesis in the Milky Way needs some other independent proof using a standard candle that is different from red clump giants but is 
of similar age.

\begin{acknowledgements}
I thank the three anonymous referees 
for comments that helped to improve the paper.
Thanks are given to Inma Mart\'\i nez-Valpuesta for some help in the interpretation of
the simulations which reproduce X-shaped pseudo-bulges, and Selcuk Bilir for providing Hipparcos
data used in his paper Bilir et al. (2008).
Thanks are given to Amy Mednick (language editor of A\&A) for proof-reading of the text.
The author was supported by the grant AYA2012-33211 of the Spanish 
Ministry of Economy and Competitiveness (MINECO). 
The VVV Survey is supported by the European Southern Observatory, by BASAL Center for Astrophysics and Associated Technologies PFB-06, by FONDAP Center for Astrophysics 15010003, by the Chilean Ministry for the Economy, Development, and Tourism's Programa Iniciativa Científica Milenio through grant P07-021-F, awarded to The Milky Way Millennium Nucleus.
\end{acknowledgements}


\begin{thebibliography}{99}

\bibitem{} Athanassoula, E. 2016, in: Galactic Bulges (Astrophys. and Space Science Library, 418), E. Laurikainen, R. Peletier, D. Gadotti, Eds., Springer International Publishing, Switzerland, p.\ 391

\bibitem{} Babusiaux, C., G\'omez, A., Hill, V., et al. 2010, A\&A,
519, A77

\bibitem{} Bensby, T., Ad\'en, T., Mel\'endez, J., et al. 2011, A\&A, 533,
A134

\bibitem{} Bensby, T., Yee, J. C., Feltzing, S., et al. 2013,
A\&A, 549, A147

\bibitem{} Bilir, S., Karaali, S., Ak, S., Yaz, E., Cabrera-Lavers, A., \& Coskunoglu, K. B.
2008, MNRAS, 390, 1569

\bibitem{} Bovy, J., Nidever, D. L., Rix, H.-W., et al. 2014, ApJ, 790, 127 

\bibitem{} Carraro, G., Ng, Y.-K., \& Portinari, L. 1998, MNRAS, 296, 1045

\bibitem{} Catchpole, R. M., Whitelock, P. A., Feast, M. W., Hughes, S. M. G.,
Irwin, M., \& Alard, C. 2016, MNRAS, 445, 2216

\bibitem{} Clarkson, W. I., Sahu, K. C., Anderson, J., et al. 2011,
ApJ, 735, 37

\bibitem{} Covey, K. R., Ivezic, Z., Schlegel, D., et al. 2007, AJ, 134, 2398

\bibitem{} D\'ek\'any, I., Minniti, D., Catelan, M., Zoccali, M., Saito, R. K.,
Hempel, M., \& Gonz\'alez, O. A. 2013, ApJ, 776, L19

\bibitem{} Girardi, L. 1999, MNRAS, 308, 818

\bibitem{} Gonz\'alez, O. A., Zoccali, M., Debattista, V. P., Alonso-Garc\'\i a, J., Valenti, E., \& Minniti, D. 2015, A\&A, 583, id. L5

\bibitem{} Gran, F., Minniti, D., Saito, R. K., et al. 2016,
arXiv:1604.01336

\bibitem{} Hewett, P. C., Warren, S. J., Leggett, S. K., \& Hodgkin, S. T. 2006,
MNRAS, 367, 454

\bibitem{} Hill, V., Lecureur, A., G\'omez, A., et al. 2011,
A\&A, 534, A80

\bibitem{} Juri\'c, M., Ivezi\'c, Z., Brooks, A., et al. 2008, ApJ, 673, 864

\bibitem{} Lee, Y. W., Joo, S.-J., \& Chung, C. 2015, MNRAS, 453, 3906

\bibitem{} L\'opez-Corredoira, M., Cabrera-Lavers, A., \& Gerhard, O. E., 2005, A\&A, 439, 107

\bibitem{} L\'opez-Corredoira, M., Cabrera-Lavers, A., Mahoney, T. J.,
Hammersley, P. L., Garz\'on, F., \& Gonz\'alez-Fern\'andez, C. 2007,
AJ, 133, 154

\bibitem{} Kunder, A., Koch, A., Rich, R. M., et al., 2012, AJ, 143, 57

\bibitem{} Mart\'\i nez-Valpuesta, I., \& Gerhard, O. E. 2013, ApJ, 766, L3

\bibitem{} McWilliam, A., \& Zoccali, M. 2010, ApJ, 724, 1491

\bibitem{} Minniti, D., Lucas, P. W., Emerson, J. P., et al. 2010, New Astron., 15, 433

\bibitem{} M\"ortsell, E. 2013, A\&A, 550, A80

\bibitem{} Nataf, D. M. 2015, arXiv:1509.00023

\bibitem{} Nataf, D. M., Udalski, A., Gould, A., Fouqu\'e, P., \& Stanek, K. Z.
2010, ApJ, 721, L28

\bibitem{} Nataf, D. M., Udalski, A., Skowron, J., et al. 2015, MNRAS, 447, 1535

\bibitem{} Ness, M., Freeman, K., Athanassoula, E., Wylie-De-Boer, E., Bland-Hawthorn, J., 
Lewis, G. F., Yong, D., Asplund, M., Lane, R. R., Kiss, L. L., \& Ibata, R. 2012, ApJ, 756,
id. 22

\bibitem{} Ness, M., \& Lang, D. 2016, AJ, 152, 14

\bibitem{} Rich R. M., 1993, in: Galaxy Evolution: The Milky Way
Perspective, ASP Conf. Ser. 49, S. R. Majewski, ed., ASP, S. Francisco,
pp. 65-82

\bibitem{} Rojas-Arriagada, A., Recio-Blanco, A., Hill, V., et al. 2014,
A\&A, 569, A103 

\bibitem{} Sadler E. M., Rich R. M., Terndrup D. M., 1996, AJ 112, 171

\bibitem{} Saito, R. K., Hempel, M., Minniti, D., et al. 2012, A\&A, 537, A107

\bibitem{} Saito, R. K., Zoccali, M., McWilliam, A., Minniti, D., Gonz\'alez, O. A., 
\& Hill, V. 2011, AJ, 142, 76

\bibitem{} Schlafly, E. F., \& Finkbeiner, D. P. 2011, ApJ, 737, 103

\bibitem{} Schlegel, D. J., Finkbeiner, D. P., \& Davis, M. 1998, ApJ, 500, 525

\bibitem{} Siegel M. H., Majewski S. R., Reid I. N., Thompson I. B.
2002, ApJ, 578, 151

\bibitem{} Tiede G. P., Frogel J. A., \& Terndrup D. M., 1995, AJ 110, 2788

\bibitem{} Wegg, C., \& Gerhard, O. 2013, MNRAS, 435, 1874

\bibitem{} Zoccali, M., Renzini, A., Ortolani, S., et al. 2003, A\&A, 399, 931

\end{thebibliography}
\end{document}